\documentclass[pdftex,twocolumn,epjc3]{svjour3}          % twocolumn

\RequirePackage[T1]{fontenc}

\smartqed  % flush right qed marks, e.g. at end of proof

\RequirePackage{graphicx}
\RequirePackage{amsmath}
\RequirePackage{multirow}
\RequirePackage{mathptmx}      % use Times fonts if available on your TeX system
\RequirePackage{flushend}
\RequirePackage[numbers,sort&compress]{natbib}
\RequirePackage[colorlinks,citecolor=blue,urlcolor=blue,linkcolor=blue]{hyperref}

\RequirePackage{verbatim}
\RequirePackage[switch,displaymath,mathlines]{lineno}
\RequirePackage[table]{xcolor}
\definecolor{lightgray}{gray}{0.9}

\newcolumntype{s}{>{\columncolor[HTML]{D7D7D7}} p{1.5cm}}
\definecolor{greenish}{RGB}{90,200,0}

\journalname{Eur. Phys. J. C}

\begin{document}

\title{Search for Double-Beta Decay of $\mathrm{^{130}Te}$ to the $0^+$ States of $\mathrm{^{130}Xe}$ with CUORE}

% author list at 02 dec 2020
\author{
D.~Q.~Adams\thanksref{USC} 
\and
C.~Alduino\thanksref{USC} 
\and
K.~Alfonso\thanksref{UCLA} 
\and
F.~T.~Avignone~III\thanksref{USC} 
\and
O.~Azzolini\thanksref{INFNLegnaro} 
\and
G.~Bari\thanksref{INFNBologna} 
\and
F.~Bellini\thanksref{Roma,INFNRoma} 
\and
G.~Benato\thanksref{LNGS} 
\and
M.~Biassoni\thanksref{INFNMiB} 
\and
A.~Branca\thanksref{Milano,INFNMiB} 
\and
C.~Brofferio\thanksref{Milano,INFNMiB} 
\and
C.~Bucci\thanksref{LNGS} 
\and
J.~Camilleri\thanksref{VirginiaTech} 
\and
A.~Caminata\thanksref{INFNGenova} 
\and
A.~Campani\thanksref{Genova,INFNGenova} 
\and
L.~Canonica\thanksref{MIT,LNGS} 
\and
X.~G.~Cao\thanksref{Fudan} 
\and
S.~Capelli\thanksref{Milano,INFNMiB} 
\and
L.~Cappelli\thanksref{LNGS,BerkeleyPhys,LBNLNucSci} 
\and
L.~Cardani\thanksref{INFNRoma} 
\and
P.~Carniti\thanksref{Milano,INFNMiB} 
\and
N.~Casali\thanksref{INFNRoma} 
\and
E.~Celi\thanksref{GSSI,LNGS} 
\and
D.~Chiesa\thanksref{Milano,INFNMiB} 
\and
M.~Clemenza\thanksref{Milano,INFNMiB} 
\and
S.~Copello\thanksref{Genova,INFNGenova} 
\and
C.~Cosmelli\thanksref{Roma,INFNRoma} 
\and
O.~Cremonesi\thanksref{INFNMiB} 
\and
R.~J.~Creswick\thanksref{USC}
% ...
} 

\institute{
Department of Physics and Astronomy, University of South Carolina, Columbia, SC 29208, USA\label{USC} 
\and
Department of Physics and Astronomy, University of California, Los Angeles, CA 90095, USA\label{UCLA} 
\and
INFN -- Laboratori Nazionali di Legnaro, Legnaro (Padova) I-35020, Italy\label{INFNLegnaro} 
\and
INFN -- Sezione di Bologna, Bologna I-40127, Italy\label{INFNBologna} 
\and
Dipartimento di Fisica, Sapienza Universit\`{a} di Roma, Roma I-00185, Italy\label{Roma} 
\and
INFN -- Sezione di Roma, Roma I-00185, Italy\label{INFNRoma} 
\and
INFN -- Laboratori Nazionali del Gran Sasso, Assergi (L'Aquila) I-67100, Italy\label{LNGS} 
\and
INFN -- Sezione di Milano Bicocca, Milano I-20126, Italy\label{INFNMiB} 
\and
Dipartimento di Fisica, Universit\`{a} di Milano-Bicocca, Milano I-20126, Italy\label{Milano} 
\and
Center for Neutrino Physics, Virginia Polytechnic Institute and State University, Blacksburg, Virginia 24061, USA\label{VirginiaTech} 
\and
INFN -- Sezione di Genova, Genova I-16146, Italy\label{INFNGenova} 
\and
Dipartimento di Fisica, Universit\`{a} di Genova, Genova I-16146, Italy\label{Genova} 
\and
Massachusetts Institute of Technology, Cambridge, MA 02139, USA\label{MIT} 
\and
Key Laboratory of Nuclear Physics and Ion-beam Application (MOE), Institute of Modern Physics, Fudan University, Shanghai 200433, China\label{Fudan} 
\and
Department of Physics, University of California, Berkeley, CA 94720, USA\label{BerkeleyPhys} 
\and
Nuclear Science Division, Lawrence Berkeley National Laboratory, Berkeley, CA 94720, USA\label{LBNLNucSci} 
\and
INFN -- Gran Sasso Science Institute, L'Aquila I-67100, Italy\label{GSSI} 
% ...
} 

\thankstext{fn1}{Deceased}

\date{Received: date / Accepted: date}

\maketitle
%\linenumbers

\begin{abstract}
The CUORE experiment is a large bolometric array searching for the lepton number violating neutrino-less double beta decay ($0\nu\beta\beta$) in the isotope 
$\mathrm{^{130}Te}$. In this work we present the latest results on two searches for the double beta decay (DBD) of $\mathrm{^{130}Te}$ to the first $0^{+}_2$ excited state of $\mathrm{^{130}Xe}$:
the $0\nu\beta\beta$ decay and the Standard Model-allowed two-neutrinos double beta decay ($2\nu\beta\beta$). Both searches are based on a 372.5 kg$\times$yr TeO$_2$ exposure.
The de-excitation gamma rays emitted by the excited Xe nucleus in the final state yield a unique signature, which can be searched for with low background by studying coincident events in two or more bolometers. The closely packed arrangement of the CUORE crystals constitutes a significant advantage in this regard.
The median limit setting sensitivities at 90\% Credible Interval (C.I.) of the given searches were \linebreak 
estimated as $\mathrm{S^{0\nu}_{1/2} = 5.6 \times 10^{24} \: \mathrm{yr}}$ for the ${0\nu\beta\beta}$ decay and \linebreak
$\mathrm{S^{2\nu}_{1/2} = 2.1 \times 10^{24} \: \mathrm{yr}}$ for the ${2\nu\beta\beta}$ decay. No significant evidence for either of the decay modes was observed and a Bayesian lower bound 
at $90\%$ C.I. on the decay half lives is obtained as: 
$\mathrm{(T_{1/2})^{0\nu}_{0^+_2} > 5.9 \times 10^{24} \: \mathrm{yr}}$ 
for the $0\nu\beta\beta$ mode and
$\mathrm{(T_{1/2})^{2\nu}_{0^+_2} > 1.3 \times 10^{24} \: \mathrm{yr}}$
for the $2\nu\beta\beta$ mode. These represent the most stringent limits on the DBD of $^{130}$Te to excited states and improve by a factor $\sim5$ the previous results on this process.
\end{abstract}
%% ----------------------------------------------------------------
\section{Introduction}
\label{sec:Intro}
%% ----------------------------------------------------------------
Double beta decay (DBD) is an extremely rare nuclear process where a simultaneous transmutation of a pair of neutrons into protons converts a nucleus (A, Z) into an isobar (A, Z+2), with the emission of two electrons and two anti-neutrinos. This two-neutrino decay mode ($2\nu\beta\beta$) is predicted in the Standard Model and was detected in several nuclei.
The neutrinoless mode of the decay ($0\nu\beta\beta$) is a posited Beyond Standard Model process that could shed light on many open aspects of modern particle physics and cosmology such as the existence of lepton number violation and elementary Majorana fermions, the neutrino mass scale, and the baryon asymmetry in the Universe \cite{Racah:1937qq,Furry:1939qr,Pontecorvo:1968,Schechter:1981bd,Fukugita:1986hr}.
Both DBD modes can proceed through transitions to the ground state as well as to various excited states of the daughter nucleus. While the former can be easier to detect through their shorter half-lives, the latter leaves a unique signature which may be detected with significantly reduced backgrounds. The excited state decays also provide powerful tests of the nuclear physics of DBD and can shed light on nuclear matrix element calculations as well as the ongoing discussion on the quenching of the effective axial coupling constant $g_A$; eventually, they could even be used to disentangle the mechanism of $0\nu\beta\beta$ decay \cite{Avignone:2007fu}.

So far, $2\nu\beta\beta$ decay to the first $0^+$ excited state has been observed in only 2 isotopes: $^{100}\mathrm{Mo}$~\cite{Barabash:1995fn} and $^{150}\mathrm{Nd}$~\cite{Barabash:2004dv}, with half lives of $(\mathrm{T}_{1/2})^{2\nu}_{0^+} =  6.1^{+1.8}_{-1.1} \times 10^{20} \: \mathrm{yr}$ and $(\mathrm{T}_{1/2})^{2\nu}_{0^+} =  1.4^{+0.4}_{-0.2} \mathrm{(stat.)} \pm 0.3 \mathrm{(syst.)} \times 10^{20} \: \mathrm{yr}$, respectively. Searches for the same process in other isotopes has yielded lower limits from $3.1 \times 10^{20} \: \mathrm{yr}$ to $8.3 \times 10^{23} \: \mathrm{yr}$ at $90 \: \% $ Confidence Level (C.L.) (see Ref.  \cite{Barabash:2017bgb} for a review).
 
In this work, we focus on the search for $0\nu\beta\beta$ and $2\nu\beta\beta$ decays of $\mathrm{^{130}Te}$ to the first $0^{+}$ excited state of $\mathrm{^{130}Xe}$ with the CUORE experiment.
Presently, the strongest limits on the decay to excited states half-life of $^{130}\mathrm{Te}$ come from a combination of Cuoricino \cite{Andreotti:2011in} and CUORE-0 \cite{Adams:2018yrj} data: 
the latter (not included in Ref. \cite{Barabash:2017bgb}) was published recently and includes the combination of the predecessor's results. The obtained limits are:
\begin{subequations}
    \begin{align}
    (\mathrm{T}_{1/2})_{0^+_2}^{0\nu} &> 1.4 \times 10^{24} \: \mathrm{yr} \; (90\%\: \mathrm{C.L.}) \label{eq:0nbbExcitedLimit_CUORE0} \\
    (\mathrm{T}_{1/2})_{0^+_2}^{2\nu} &> 2.5 \times 10^{23} \: \mathrm{yr} \; (90\%\: \mathrm{C.L.}) \label{eq:2nbbExcitedLimit_CUORE0} 
    \end{align}
\end{subequations}
for the $0\nu$ and $2\nu$ process, respectively.
Theoretical predictions~\cite{PhysRevC.91.054309} on the $\mathrm{ (T_{1/2}) }^{2\nu}_{0^+_2}$ observable in the $2\nu\beta\beta$ decay channel are based on the QRPA approach and favor the following range: 
\begin{equation}
    ^{th}(\mathrm{T}_{1/2})^{2\nu}_{0^+_2} = (7.2 - 16) \times 10^{24} \: \mathrm{yr}
    \label{eq:2nbbExcitedTheory}
\end{equation}
where the range depends on the precise treatment of $g_A$. The lower bound assumes a constant function of the mass number A, and the upper bound assumes a value of $g_A = 0.6$ \cite{PhysRevC.91.054309,SUHONEN2013153,SUHONEN20141}.

A data driven estimate of the $2\nu\beta\beta$ ground state to excited state decay rate in the IBM-II framework based on Ref. \cite{Barea:2013bz,Kotila:2012zza,Barabash:2010ie} is reported in Ref. \cite{Lehnert2015} as
\begin{equation}
    ^{th}( \mathrm{T}_{1/2})^{2\nu}_{0^+_2} = 2.2 \times 10^{25} \: \mathrm{yr}.
\end{equation}
In this regard, as stated before, both a measurement or a more stringent limit with respect to Ref. \cite{Adams:2018yrj} are informative from the point of view of refining and validating the theoretical computations. 

The decay to excited states has a unique signature. The double-beta decay emits two electrons, which share kinetic energy up to 734 keV. The subsequent decay of the excited daughter nucleus typically emits two or three high energy gamma rays in cascade. Due to the emission of such coincident de-excitation $\gamma$ rays, both $0\nu\beta\beta$ and $2\nu\beta\beta$ decay channels allow a significant background reduction with respect to the corresponding transitions to the ground state. This holds especially in an experimental setup that exploits a high detector granularity, such as the CUORE experiment. 

%% ----------------------------------------------------------------
\section{Detector and Data Production}
\label{sec:DetectorDataProduction}
%% ----------------------------------------------------------------
The Cryogenic Underground Observatory for Rare Events (CUORE)  \cite{Arnaboldi:2002du,Brofferio:2019yoc} is a ton-scale cryogenic detector located at the underground Laboratori Nazionali del Gran Sasso \linebreak (LNGS) in Italy. 
CUORE is designed to search for the $0\nu\beta\beta$ decay of $\mathrm{^{130}Te}$ to the ground state of $\mathrm{^{130}Xe}$ \cite{Alduino:2017ehq,Adams:2019jhp}, and has a low background rate near the $0\nu\beta\beta$ decay region of interest (ROI), an excellent energy resolution, and a high detection efficiency.
The CUORE detector consists of a close-packed array of 988 TeO$_2$ crystals operating as cryogenic bolometers \cite{Fiorini:1983yj,Enss:2008x,Arnaboldi:2010fj} at a temperature of $\sim$10 mK. \linebreak
The CUORE crystals are $5 \times 5 \times 5$ cm$^3$
cubes weighing 750 g each, arranged in 19 towers: each consisting of a copper structure with 13 floors and 4 crystals per floor.
A custom-made $^3$He/$^4$He dilution refrigerator, which represents the \linebreak
state of the art for this cryogenic technique, is used to cool down the CUORE cryostat, where the entire array is contained and shielded.   \cite{Alduino:2019xia,Alessandria:2013ufa,Buccheri:2014bma,Arnaboldi:2017aek,DiDomizio:2018ldc,Benato:2017kdf,DAddabbo:2017efe}. 

Each CUORE crystal records thermal pulses via a \linebreak neutron-transmutation doped (NTD) germanium thermistor \cite{Haller:1984ntd} glued to its surface. Any energy deposition in the crystal causes a sudden rise in temperature and can indicate the emission of a particle inside, the crossing of a particle \linebreak 
through, or some environmental thermal instability (e.g. \linebreak 
earthquakes).

The data acquisition and production of CUORE event data used in this work closely follows the procedure used in \cite{Alduino:2019xia} and is described in detail in \cite{Alduino:2016zrl}. We briefly review the basic process here and highlight the differences.

The NTD converts the thermal signal to a voltage output, which is amplified, filtered through a 6-pole Bessel anti-aliasing filter, and sampled continuously at 1kHz. The data are stored to disk and triggered offline with an algorithm based on the optimum filter (OF) \cite{Gatti:1986cw,DiDomizio:2010ph,campani2020lowering}.

For each triggered pulse, a 10 second window around each trigger (3 seconds before and 7 seconds after) is processed through a series of analysis steps, with the aim of extracting the physical quantities associated to the pulse.
The waveform is filtered using an OF built from a pulse template, and the measured noise power spectrum.

The signal amplitudes are then evaluated from the OF filtered waveforms and those amplitudes are corrected for small changes in the detector gain due to temperature drifts. 
We calibrate each bolometer individually using dedicated calibration runs with $\mathrm{^{232}Th}$ and $\mathrm{^{60}Co}$ gamma sources deployed around the detector array. These calibration runs typically last a few days every two months.

We impose a pulse shape selection (PSA) based on 6 pulse shape parameters. This cut removes noisy events, \linebreak 
pileup events, and non-physical events.

Unlike the decay to ground state search described in Ref. \cite{Alduino:2019xia}, the physics search described in the present work focuses on coincident energy depositions in multiple crystals. In particular, we are focusing on events where energy is deposited in either two or three bolometers.
As the reconstructed time difference between events on nearby bolometers is affected by differences in pulse rise times, a bolometer-by-bolometer correction is applied.
Sets of coincident energy releases in $M$ bolometers within a $\pm 5\:$~ms time window are grouped together as multiplets of multiplicity $M$.

CUORE started its data taking in May 2017 and, after two significant interruptions for important maintenance of the cryogenic system, is now seeing its exposure grow at an average rate of $\sim$ 50 kg$\times$yr/month.
The CUORE data collection is organized in datasets: a dataset begins with a gamma calibration campaign that typically lasts 2-3 days, followed by 6-8 weeks of uninterrupted background data taking, and ends with another gamma calibration.

Recently, the CUORE collaboration released the results of the search for $0\nu\beta\beta$ decay to the g.s. on the the accumulated exposure of 372.5 kg$\cdot$y, setting an improved limit on the half-life of $\mathrm{^{130}Te}$ of $(\rm{T_{1/2}})^{0\nu}_{0^+_1} > 3.2\times10^{25}$ yr \cite{Adams:2019jhp}.

%% ----------------------------------------------------------------
\section{Analysis}
\label{sec:Analysis}
\begin{figure}
    \centering
    \includegraphics[width=0.5\textwidth]{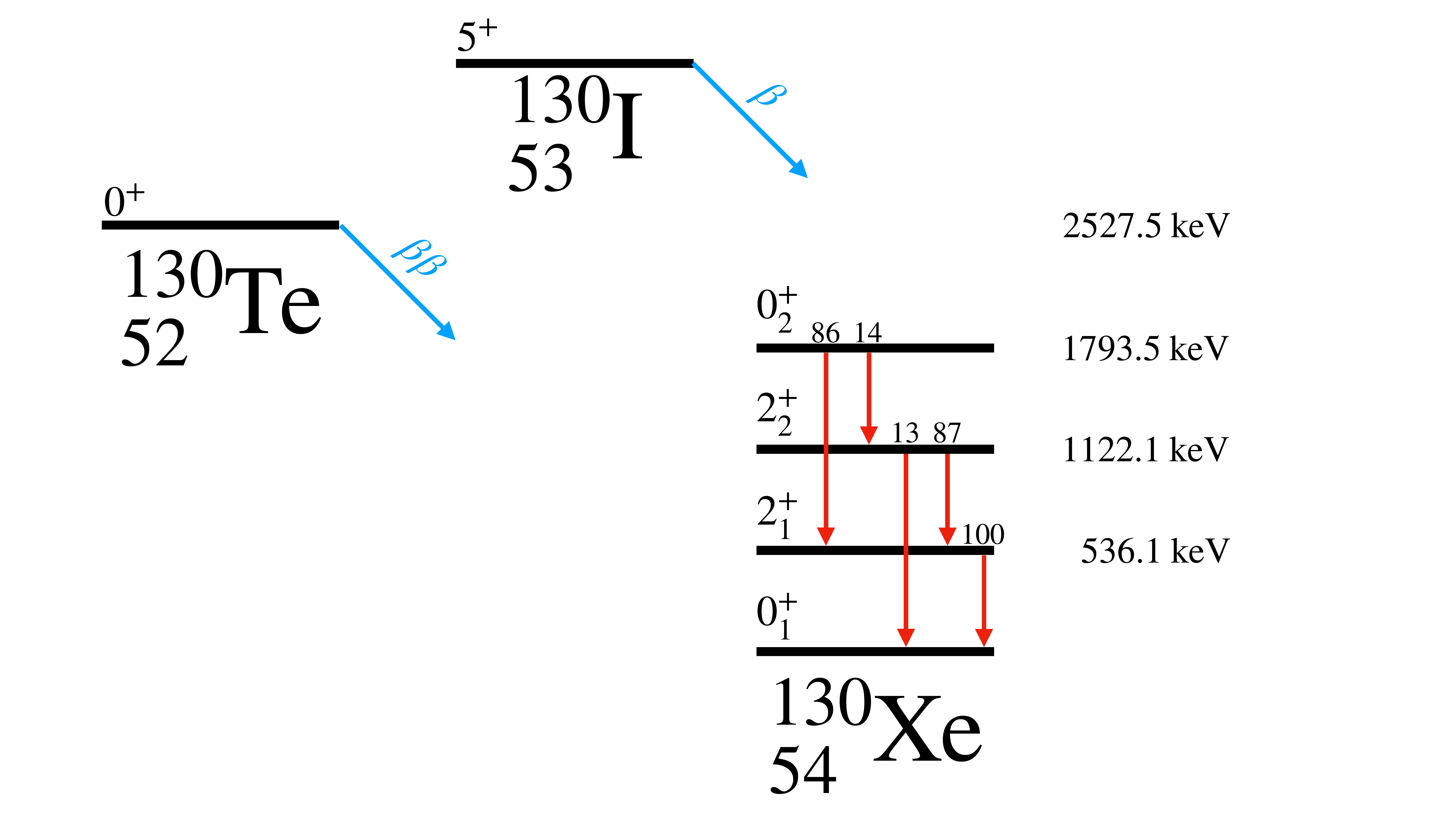}
    \caption[$^{130}\mathrm{Te}$ $\beta\beta$ decay scheme to excited states ]
    {The decay scheme of $^{130}\mathrm{Te}$ is shown with details about the involved excited states of $^{130}\mathrm{Xe}$ up to its first $0^+$ excited state. The nomenclature $0^+_1, ..., 0^+_n$ indicates states with the same angular momentum in increasing order of excitation energy. An energy scale is shown (right) where the $^{130}\mathrm{Xe}$ ground state is taken as reference \cite{SINGH200133}.}
    \label{fig:ExcitedStatesDecayScheme}
\end{figure}

%% ----------------------------------------------------------------
In this section we describe the analysis steps that are specific to the search for $^{130}\mathrm{Te}$ decay to the excited states of $^{130}\mathrm{Xe}$.
The de-excitation of the $^{130}\mathrm{Xe}$ nucleus follows one of three possible \textit{patterns}, i.e. paths through states of decreasing energy from the $0^+_2$ to the $0^+_1$ ground state (Figure \ref{fig:ExcitedStatesDecayScheme}). Details about the probability of each de-excitation pattern, referred in the following as A, B and C (in decreasing order of probability), and the energy of the emitted $\gamma$ rays are reported in Table \ref{tab:DeExcitationPatterns}.

The simultaneous emission of DBD betas and \linebreak
de-excitation gammas produces coincidence multiplets, i.e. sets of simultaneous pulses in $M$ bolometers, grouped by the coincidence algorithm. We search for events with \textit{full containment} of the final state gammas in the crystals: more specifically we try to avoid multiplets where one or more of the final state $\gamma$s escape the source crystal and are absorbed by some non-active part of the experimental apparatus, or Compton scattering events, where the energy of a single de-excitation gamma is split among two or more detectors. We place energy selection cuts to find these events, which are listed in Table \ref{tab:SelectedScenarioES} and described in more details in Sec. \ref{sec:Ranking}.

\textit{Partitions} are defined as unique groupings of energy depositions that pass a particular set of energy selection cuts. For a fixed multiplicity $M$ and a source pattern, they are identified by all possible ways of partitioning the final state particles in $M$ different crystals. Finally we define \textit{signatures} as partitions from different patterns that are indistinguishable. Single-site ($M = 1$) signatures are not taken into account, as the $0\nu\beta\beta$ decay channel would be indistinguishable from the same decay on the ground state of $^{130}\mathrm{Xe}$, while the $2\nu\beta\beta$ decay channel, instead, would suffer from high background from the decay to the ground state. 
Therefore, there remain $8$ partitions for patterns A and B, and $14$ for pattern C. Each of the partitions is labelled with strings of $3$ characters with the following convention
\begin{center}
    [\textit{Multiplicity}] [\textit{Pattern}] [\textit{Index}]
\end{center}
where \textit{Multiplicity} = 2,3,4 indicates the number of involved crystals, \textit{Pattern} = A,B,C stores the originating de-excitation pattern, and \textit{Index} is a unique integer counter to distinguish the various combinations of energy groupings for that pattern and multiplicity. Partitions sharing the same expected energy release are indistinguishable and are merged as signatures. For this reason, instead of handling a total number of $22$ partitions we are left with $15$ signatures \cite{excitedTAUP19}.
An example of indistinguishable partitions is given in Table \ref{tab:SelectedScenarioES} by the 2A0-2B1 signature. In one of the crystals a gamma energy release of 1257~keV is expected. This can be either due (see Table \ref{tab:DeExcitationPatterns}) to $\gamma_1$ from pattern A or the simultaneous absorption of $\gamma_1 + \gamma_2$ from pattern B. 

%% ------------------------------- 

\begin{table}[]
    \centering
    \begin{tabular}{ccccc}
        \textbf{Pattern}    & \textbf{BR [\%]}  &   \textbf{Energy $\gamma_1$}  &   \textbf{Energy $\gamma_2$}  &   \textbf{Energy $\gamma_3$} \\
        \hline
         A & $86\%$ & $1257 \: \mathrm{keV}$    & $536 \: \mathrm{keV}$     &   -  \\
         B & $12\%$ & $\:\:671 \: \mathrm{keV}$    & $ 586 \: \mathrm{keV}$    & $ 536 \: \mathrm{keV}$ \\
         C & $2\%$  & $1122 \: \mathrm{keV}$    & $ 671 \: \mathrm{keV}$    &   - \\
    \end{tabular}
    \caption[De-excitation patterns of the $0^+_2$ state of $^{130}\mathrm{Xe}$]
    {The de-excitation $\gamma$ rays emitted by $^{130}\mathrm{Xe}^*$ in the transition from the $0^+_2$ to the ground state. Each row corresponds to a different path through intermediate states. The energies of the emitted $\gamma$s are listed, in order of energy, along with the branching ratio (BR) of each pattern \cite{SINGH200133}.}
    \label{tab:DeExcitationPatterns}
\end{table}

%% ----------------------------------------------------------------
\subsection{Monte Carlo Simulations}
\label{sec:MonteCarloSimulations}
%% ----------------------------------------------------------------
We use Monte Carlo (MC) simulations to compute the detection efficiency (Sec. \ref{sec:EfficiencyEvaluation}) and the expected background (Sec. \ref{sec:Background}) associated with each signature, to rank experimental signatures and eventually fine tune selection cuts on the most sensitive ones (Sec. \ref{sec:Ranking}).   

CUORE uses a Geant4-based MC to simulate energy depositions in the detector. The Geant4 software \cite{G4_AGOSTINELLI2003250} simulates particle interactions in the various volumes and materials of a modeled detector geometry. A separate post-processing step converts the resulting energy depositions into an output as close to the output of the data production as possible.
We refer to Ref.~\cite{Alduino_2017} for further details about the CUORE MC simulations.

Signal simulations, that are simulations of the double beta decay to excited states and the subsequent de-excitation gammas, are produced separately for each process ($0\nu$,$2\nu$) and pattern (A,B,C). 
Gamma energies are generated as \linebreak 
monochromatic. 
Angular correlations induce a negligible effect on the containment efficiency of the experimental signatures listed in Table \ref{tab:SelectedScenarioES} as opposed to isotropic gamma emission and compared with the dominant systematic uncertainty described in Sec. \ref{sec:SystematicUncertainties}.
Beta energies are randomly extracted from the beta spectrum of the corresponding decay \cite{Haxton:1985am,Doi:1981mi,Doi:1981mj,Kotila:2012zza} in the HSD hypothesis \footnote{The $2\nu\beta\beta$ is called single state
dominated (SSD) if it is governed by the lowest $1^+$ energy level. In the higher-state dominated (HSD) case the calculation is simplified by summing over all the virtual intermediate states and assuming an average closure energy.
In the SSD hypothesis the cumulative sum energy distribution of emitted electrons \cite{Kotila:2012zza} differs by $<0.3\%$ with respect to HSD.
We note though that this analysis cannot infer the shape of the beta spectrum because in $2\nu\beta\beta$ signatures the fit is performed just on the gamma peaks.}

Background simulations take as input the CUORE background model~\cite{CUORE2nuBB}, and include contaminations in the crystal and several other parts of the CUORE setup\footnote{The CUORE background model we refer to is still preliminary, however the estimates of the background activities are good enough to understand what will be the expected contribution to the present search. In the final fit the exact values are floated.}, such as: the copper tower structure, the closest copper vessel enclosing the detector, the Roman lead, the internal and external modern lead shields and the internal lead suspension system. The contaminants include bulk and surface $^{238}$U and $^{232}$Th chains with different hypotheses on secular equilibrium breaks, bulk $^{60}$Co, $^{40}$K, and a few other long lived isotopes. Additional sources of background included are the cosmic muon flux and the $2\nu\beta\beta$ decay of $^{130}$Te to the ground state. 
Both signal and background simulated energy spectra are convolved with a Gaussian resolution that has a width of $5~\mathrm{keV}$ full width at half maximum, a standard choice for our simulation studies~\cite{Alduino_2017}.

%% ----------------------------------------------------------------
\subsection{Efficiency Evaluation}
\label{sec:EfficiencyEvaluation}
%% ----------------------------------------------------------------
 
The detection efficiency of a given signature consists of two components: the containment efficiency and the analysis efficiency. Given a signature $s$ and a set of energy selection cuts on the involved bolometers, the corresponding containment efficiency $\varepsilon_s$ represents the probability that the energy released by a nuclear decay of $^{130}\mathrm{Te}$ to the $0^+_2$ state of $^{130}\mathrm{Xe}$ matches the topology of the signature. We evaluate this efficiency component from the signal MC simulations described in Sec. \ref{sec:MonteCarloSimulations}, by summing over the contributions of all patterns $p$ populating the signature $s$
\begin{equation}
    \varepsilon_s = 
    \Bigg[ \sum_{p} \mathrm{BR}_p \cdot \frac{ \big[ N^{(sel)}_{MC} \big]^{(s)}_p }{ \big[N^{(tot)}_{MC}\big]_p\:\:\: }  \Bigg]
    \label{eq:ContainmentEfficiency}
\end{equation}
where $\mathrm{BR}_p$ is the branching ratio of pattern $p$, $\big[ N^{(sel)}_{MC} \big]^{(s)}_p$ and $\big[ N^{(tot)}_{MC} \big]_p$ are respectively the selected and total number of simulated decays in the de-excitation pattern of interest.
For $0\nu\beta\beta$ decay signatures the signal is monochromatic in all the involved crystals, so the signal region is expected to lie around a specific point in the M-dimensional space of coincident energy releases. A selection is enforced, in simulations, with a \textit{box cut}, i.e. a selection interval for energy releases in each crystal, defined as
\begin{equation}
    | E_i - Q_i | < 5 \: \mathrm{keV} \quad \mathrm{where} \quad i = 1 ... M
    \label{eq:SelectionPeak}
\end{equation}
where $E_i$ is the reconstructed energy release in the ordered energy space\footnote{The energy releases of each \texttt{M}-bolometers multiplet are ordered in descending order so that $E_i > E_{i+1}$.} and $Q_i$ is the corresponding expected energy release.
For $2\nu\beta\beta$ decay signatures the same selections apply except the one crystal where the energy release from the $\beta\beta$ is expected.
Since the emitted neutrinos carry away an unknown (on an event basis) amount of undetected  energy, the expected energy release is not monochromatic. It is instead expected to vary from $Q_{j}^{min}$ to $Q_{j}^{max}$ where $j$ indicates the bolometer where the $\beta\beta$ release their energy. For that bolometer, in each multiplet the following selection is applied
\begin{equation}
     Q_{j}^{min} - 5 \: \mathrm{keV} < E_j < Q_{j}^{max} + 5 \: \mathrm{keV}
     \label{eq:SelecionContinuum}
\end{equation}

Selection cuts need to be further tuned at a later stage to optimize the sensitivity to signal peaks. We do this including the widest possible sidebands around each signal peak in order to best constrain the underlying continuous background. We try to avoid including background peaks in the fit range, in order to minimize systematics due to their modeling.
This process yields the selections listed in Table \ref{tab:SelectedScenarioES} (see Sec. \ref{sec:Ranking}). We then update our computation of signal efficiencies using Eq. \ref{eq:ContainmentEfficiency}, where $N^{sel}_{MC}$ is replaced by the result of a Gaussian fit to the distribution of selected MC signal events.

The second efficiency contribution, namely the analysis efficiency, is the combination of the probability of correctly detecting and reconstructing the energy deposited in each bolometer (cut efficiency, $\varepsilon^{cut}$), and the probability of assigning the correct multiplicity and avoiding an accidental coincidence (accidentals efficiency, $\varepsilon^{acc}$).

The cut efficiency term is named after the data processing cuts needed to select triggered events that pass the base and PSA cuts (see Sec. \ref{sec:DetectorDataProduction}). The method used to calculate this efficiency follows closely what was used in \cite{Adams:2019jhp}.
We measure the efficiency of correctly triggering, reconstructing the pulse energy and the pile-up contribution (base cuts) from heater pulses. The base cut efficiency is computed on each bolometer-dataset pair given the large number of available heater events, and then averaged to obtain a per-dataset value. 
The PSA cut efficiency is extracted from two independent samples of events: either coincident double-site events where the total energy released is compatible with known prominent $\gamma$ lines, or single-site events due to fully absorbed $\gamma$ lines. The first sample includes events whose energy spans a wide range, and allows the determination of the PSA cut efficiency dependence on energy. The second sample has a higher statistics but provides a measurement at the energies of the selected $\gamma$ peaks only, rather than on a continuum.
We evaluate for each dataset the PSA cut efficiency term as the average of the two efficiencies obtained from such samples. We treat the difference as a systematic effect.
The $\varepsilon^{cut}$ term must be raised to the $M^{th}$ power because it models bolometer-related efficiencies and a multiplet is selected if and only if all of the involved bolometers pass the selection cuts. 

The accidentals efficiency term $\varepsilon^{acc}$ is obtained, separately for each dataset, as the survival probability of the $^{40}$K $\gamma$ line at 1460~keV. A fully absorbed $^{40}$K line in CUORE is uncorrelated to any other physical event because it follows an electron capture of a $\sim3$~keV shell, which is below threshold.

Summarizing, the total signal detection efficiency for signature $s$ is
\begin{equation}
    \varepsilon^{tot} =  
        \varepsilon_s\times
        (\varepsilon^{cut})^M \times \varepsilon^{acc}.
    \label{eq:Efficiency}
\end{equation}
Since the cut efficiency and accidentals term are evaluated separately for each dataset, the total efficiency term in Eq. \ref{eq:Efficiency} must be thought of as the signal efficiency for signature $s$ for a specific dataset. A summary of the relevant efficiency values is provided in Table \ref{tab:Efficiency}, where the per-dataset values are exposure weighted over all datasets. The containment efficiency is the dominant term.

%% old table layout before CollaborationReview
\begin{comment}
\begin{table}[]
    \centering
    \begin{tabular}{lccc}
                        & \textbf{2A0-2B1}  & \textbf{2A2-2B3}  & \textbf{3A0}  \\
    \hline
    Containment         & $4.2(2) $  & $2.4(1) $  & $0.19(1)$ \\
    Cut                 & \multicolumn{2}{c}{$78.7(2)$}           & $69.8(3)$ \\
    Accidentals         & \multicolumn{3}{c}{$98.7(1)$} \\
    \hline
    Total               & $3.2(1) $  & $1.9(1)$  & $0.13(1)$ \\
    \end{tabular}
    \caption{ We report the efficiency terms that appear in Eq. \ref{eq:Efficiency} separately for the $2\nu\beta\beta$ (top) and $0\nu\beta\beta$ (bottom) analyses. The containment term dominates the efficiency. We report the cut efficiency raised to power $M$ according to the signature it refers to. We quote effective values computed as exposure weighted mean for the cut and accidentals efficiency terms. All values are percentages, the uncertainty on the last digit is included in round brackets.
    }
    \label{tab:Efficiency}
    
    \begin{tabular}{lccc}
                        & \textbf{2A0-2B1}  & \textbf{2A2-2B3}  & \textbf{3A0}  \\
    \hline
    Containment         & $4.6(2)$  & $2.9(1)$  & $2.5(1)$ \\
    Cut                 & \multicolumn{2}{c}{$78.7(2)$}           & $69.8(3)$ \\
    Accidentals         & \multicolumn{3}{c}{$98.7(1)$} \\
    \hline
    Total               & $3.5(2)$  & $2.3(1)$  & $1.7(1)$ \\
    \end{tabular}
\end{table}
\end{comment}

%% new table layout: swapped 0nbb and 2nbb (top/bottom) for consistency with the following tables
\begin{table}[]
    \caption{ We report the efficiency terms that appear in Eq. \ref{eq:Efficiency} separately for the $0\nu\beta\beta$ (top) and $2\nu\beta\beta$ (bottom) analyses. The containment term dominates the efficiency. We report the cut efficiency raised to power $M$ according to the signature it refers to. We quote effective values computed as exposure weighted mean for the cut and accidentals efficiency terms. All values are percentages, the uncertainty on the last digit is included in round brackets.
    }
    \label{tab:Efficiency}
    
    \centering
    \begin{tabular}{lccc}
    \multicolumn{4}{c}{$0\nu\beta\beta$} \\
    \hline
                        & \textbf{2A0-2B1}  & \textbf{2A1-2B2}  & \textbf{3A0}  \\
    Containment         & $4.6(2)$  & $2.9(1)$  & $2.5(1)$ \\
    Cut                 & \multicolumn{2}{c}{$78.7(2)$}           & $69.8(3)$ \\
    Accidentals         & \multicolumn{3}{c}{$98.7(1)$} \\
    Total               & $3.5(2)$  & $2.3(1)$  & $1.7(1)$ \\
    \\
    \end{tabular}
    
    \begin{tabular}{lccc}
    \multicolumn{4}{c}{$2\nu\beta\beta$} \\
    \hline
                        & \textbf{2A0-2B1}  & \textbf{2A1-2B2}  & \textbf{3A0}  \\
    Containment         & $4.2(2) $  & $2.4(1) $  & $0.19(1)$ \\
    Cut                 & \multicolumn{2}{c}{$78.7(2)$}           & $69.8(3)$ \\
    Accidentals         & \multicolumn{3}{c}{$98.7(1)$} \\
    Total               & $3.2(1) $  & $1.9(1)$  & $0.13(1)$ \\
    \end{tabular}

\end{table}

\subsection{Background Contributions}
\label{sec:Background}
Radioactive decays and particle interactions other than $\mathrm{^{130}Te}$ decay to $\mathrm{^{130}Xe}$ excited state, may mimic the process we search for. We estimate this background contribution by \linebreak
means of background MC simulations described in Sec \ref{sec:MonteCarloSimulations}. We combine background simulations of different sources, according to the CUORE background model, and from the simulated background spectra we compute the expected number of background counts for each signature $B_s$, by counting the expected events from each source included in the background model, and summing the contributions from all sources. We apply the same tight selection cuts around the signal region defined in Eqs. \ref{eq:SelectionPeak} and \ref{eq:SelecionContinuum}.

We use $B_s$ to evaluate an approximate sensitivity for each signature and ultimately select the ones that will enter the analysis (see Sec. \ref{sec:Ranking}).
Once the signatures that enter the analysis are selected, we optimize the selection cuts around the signal region in order to reject background structures while leaving the widest possible sidebands around the expected signal position. In this way we can parameterize the background with an appropriate analytical function, whose shape is dictated by background simulations, and use that to perform the final analysis (see Sec. \ref{sec:fitting}). With this method we infer the number of reconstructed background events in each signature from data, rather than relying just on simulations.

\subsection{Experimental Signature Ranking}
\label{sec:Ranking}

The 15 unique signatures under analysis have different signal efficiencies and backgrounds, and thus different detection sensitivities of the signal. In this section we evaluate an approximate sensitivity of each signature and reduce the 15 signatures down to the most sensitive subset.

We analytically evaluate the discovery sensitivity of signature $s$ starting from a background-only model for the total number of counts observed in a single bin centered at the expected signal position. In background-free signatures $B_s \ll 1$ \linebreak
we assume an exponentially decaying prior P$(\mu) = e^{-\mu}$ \linebreak 
where $\mu$ is the true value of the number of background \linebreak
counts. In background-limited ones $B_s \gg 1$ we assume a Gaussian prior whose mean and variance are $B_s$. We define the discovery sensitivity as the minimum number of observed counts $N_s$ such that the probability of observing $N > N_s$ counts in the background-only model is smaller than a given threshold $p_{th}$. Then, from $N_s$, we extract the corresponding half life sensitivity
\begin{equation}
    \Tilde{S}_{1/2}(\varepsilon_s,B_s) = 
    \bigg[ \frac{ \ln(2) \: M\Delta t \: N_A \: \eta(^{130}\mathrm{Te})}{m(\mathrm{TeO}_2)} \bigg] S(\varepsilon_s,B_s)
    \label{eq:SensitivityTilde}
\end{equation}
where $M$ is the detector mass, $\Delta t$ its live time, $N_A$ the Avogadro constant, $\eta(^{130}\rm{Te}) = (34.167 \pm 0.002) \%$ \cite{Fehr:2004} the isotopic abundance of $^{130}\mathrm{Te}$ in natural tellurium, \linebreak $m(\mathrm{TeO}_2) = 159.6$~g/mol the molecular mass of a tellurium dioxide molecule \cite{Fehr:2004} and $S(\varepsilon_s,B_s)$ is a score function
\begin{equation}
    S(\varepsilon_s,B_s) = 
    \begin{cases}
        \frac{\varepsilon_s}{-\ln(p_{th})} & B_s < B_{th} \\
        \frac{\varepsilon_s}{n_{\sigma}(p_{th})\sqrt B_s} & B_s \geq B_{th}
    \end{cases}
\end{equation}
where $n_{\sigma}(p_{th})$ is the number of Gaussian sigma which correspond to $p_{th}$, and $B_{th}$ sets the transition from the \linebreak 
background-free approximation to the background-limited approximation making $S(\varepsilon_s,B_s)$ continuous.
For $n_\sigma = 5$, \linebreak
$p_{th} \sim 3 \times 10^{-7}$ and $B_{th} \sim 9$. We note though that all signatures have a number of expected background counts either $<1$ or $>10$ and their ranking would not be affected by a different choice of the $B_{th}$ threshold.

We compute the relative score of each signature $s$ as 
\begin{equation}
    R_s \doteq \frac{S(\varepsilon_s,B_s)}{\sum_{s'} S(\varepsilon_{s'},B_{s'})}
    \label{eq:RelativeScore}
\end{equation}
where $s'$ is an index running on all experimental signatures. 
The efficiency term $\epsilon_s$ only includes the MC-based containment efficiency term. This is acceptable for the computation of an approximate analytical score function, since the containment term by far dominates the overall efficiency (Tab. \ref{tab:Efficiency}). 

We set a threshold of $R_s > 5 \%$ and we identify three signatures both for the $0\nu\beta\beta$ and $2\nu\beta\beta$ decay search, namely the 2A0-2B1, 2A2-2B3 and 3A0 listed in Table \ref{tab:SelectedScenarioES}. The selected experimental signatures account for a majority of the sensitivity contributions among studied signatures. The sum of their scores accounts for $84 \% \: (87 \%)$ total score of all signatures in the $0\nu\beta\beta \: (2\nu\beta\beta)$ search respectively.

\begin{table}[]
\centering

    \caption[]{Selected experimental signatures for DBD search on the $0^+_2$ excited state of $^{130}\mathrm{Xe}$ in the $0\nu\beta\beta$ (top) and $2\nu\beta\beta$ (bottom) channel are listed. For each signature the corresponding Regions Of Interest (ROI, i.e. the applied selection cuts) are listed in terms of the ordered energy releases $E_{1} \geq E_{2} \geq E_{3}$. The component that will be used for the fit is highlighted with a $^*$ superscript. For each signature the partition of the secondaries expected to contribute are listed. For each secondary we report in round brackets the expected energy release in keV. The last row reports the corresponding relative score (Eq. \ref{eq:RelativeScore}). }
    
    \label{tab:SelectedScenarioES}

\begin{tabular}{ p{2.4cm} p{2.4cm} p{2.4cm} }
%\begin{tabular}{|c|c|c|}
    \multicolumn{3}{c}{$0\nu\beta\beta$} \\
    \hline
    \textbf{2A0-2B1} & \textbf{2A2-2B3} & \textbf{3A0} \\
    \hline
    $E_{2}: \gamma (1257)$                      &  $E_{2}: \gamma (536)$    &   $E_{3}: \gamma (536)$ \\
    $E_{1}: \beta\beta (734) + \gamma  (536)$   &  $E_{1}: \beta\beta (734) + \gamma (1257)$ &  $E_{2}: \beta\beta (734)$ \\
                & &   $E_{1}: \gamma (1257)$ \\

    \hline
    $1247 < E_{2} < 1280 $      & $523 < E^*_{2} < 573 $   & $526 < E_{3} < 546 $\\
    $1247 < E_{1}^* < 1280 $    & $1981 < E_{1} < 2001 $   & $700 < E^*_{2} < 760 $ \\
                                                        & & $1247 < E_{1} < 1267 $ \\
    \hline
    \textit{$39 \%$} & \textit{$25 \%$} & \textit{$20 \%$} \\
    \hline
    \\
\end{tabular}

\begin{tabular}{p{2.4cm} p{2.4cm} p{2.4cm} }
%\begin{tabular}{|c|c|c|c|}
    \multicolumn{3}{c}{$2\nu\beta\beta$} \\
    \hline
    \textbf{2A0-2B1} & \textbf{2A2-2B3} & \textbf{3A0} \\
    \hline

     $E_{2}: \beta\beta (0-734) + \gamma (536)$ &  $E_{2}: \gamma (536)$                        &  $E_{3}: \beta\beta (0-734)$ \\
     $E_{1}: \gamma (1257)$                     &  $E_{1}: \beta\beta (0-734) + \gamma (1257)$  &  $E_{2}: \gamma (536)$ \\
                                                                                                & &  $E_{1}: \gamma (1257)$ \\
\hline
    $620 < E_{2} < 1150 $       &  $523 < E^*_{2} < 573 $   &  $400 < E_{3} < 523 $\\
    $ 1220 < E^*_{1} < 1300 $   &  $1360 < E_{1} < 1990 $   &  $523 < E^*_{2} < 573 $ \\
                                                            & &  $1779 < E_{1}+E_{2} < 1807 $ \\
    \hline
    \textit{$40 \%$} & \textit{$22 \%$} & \textit{$25 \%$} \\
    \hline
\end{tabular}

\end{table}

%% ----------------------------------------------------------------
\section{Physics Extraction}
%% ----------------------------------------------------------------
We use a phenomenological parameterization of the background in the fitting regions (as opposed to using the predicted spectra from the MC), hence real data are required to tune the fit. To avoid biasing our results, we build the fit (Sec. \ref{sec:fitting}) on blinded data using the blinding procedure described in Sec. \ref{sec:BlindingAndSensitivity}.

%% ----------------------------------------------------------------
\subsection{Fitting Technique}
\label{sec:fitting}
%% ----------------------------------------------------------------
We extract the $0\nu\beta\beta$ decay rate and $2\nu\beta\beta$ decay rate of $^{130}\mathrm{Te}$ to the $0^+_2$ excited state of $^{130}\mathrm{Xe}$ using two separate Bayesian fits. For the single process ($0\nu$ or $2\nu$) the fit is run simultaneously on all the involved signatures.
Every multiplet of multiplicity $M$ can be represented as a point $\vec E_{ev}$ in a $M$-dimensional space of reconstructed energies. The energy releases are ordered so that $E_i > E_{i+1} \:\: \forall i = 1, .. ,M-1$. For each signature, one of the components of the $\vec E_{ev}$ vector is selected to perform the fit. This is referred to as \textit{projected energy}, and indicated with a $^{*}$ superscript in Table \ref{tab:SelectedScenarioES}. In the following we will denote this energy as $E_{ev}$.

An unbinned Bayesian fit is implemented with the BAT software package \cite{Caldwell:2008fw}. It allows simultaneous sampling and maximization of the posterior probability density function (pdf) via Markov Chain Monte Carlo. The likelihood can be decomposed, for each signature and dataset, as follows:
\begin{equation}
\begin{split}
    \log \mathcal L_{s,ds} & = - ( \lambda_{S \: s,ds} + \lambda_{B \: s,ds} ) + \\
    & + \sum_{ev \in (s,ds)} \log \bigg[ \lambda_{S \: s,ds} 
    \xi_{bo,ds} f_S( E_{ev} ) + \\
    & + \lambda_{B \: s,ds} \xi_{bo,ds} f_B( E_{ev} ) \bigg]
\end{split}
\label{eq:Likelihood_SDS}
\end{equation}
where the subscripts $s$, $bo$, $ds$ will be used to refer to a specific signature, bolometer, or dataset respectively. The form of Eq. \ref{eq:Likelihood_SDS} is the same for $0\nu\beta\beta$ and $2\nu\beta\beta$,
the $\lambda_S$ and $\lambda_B$ terms are the expected number of signal and background events respectively, $\xi_{bo,ds}$ is the ratio between the exposure of bolometer $bo$ to the exposure of dataset $ds$, $f_S$ and $f_B$ are the normalized signal and background pdfs. 
They depend just on the projected energy variable $E_{ev}$.

The response function of CUORE bolometers to \linebreak
monochromatic energy releases has a functional form defined phenomenologically for each bolometer-dataset pair \cite{Alexey2018} \cite{Laura2018} as the superposition of 3 Gaussian components to account for non-Gaussian tails. A correction for the bias in the energy scale reconstruction is implemented together with the resolution dependence on energy (see Ref.~\cite{Alduino:2017ehq} for more details). The signal term $f_S(E_{ev})$ models such shape in the bolometer-dataset pair the projected energy $E_{ev}$ was released in. 
The expected number of signal counts can be written as
\begin{equation}
\begin{split}
 \lambda_{S \: s,ds} & = 
                        \Gamma^{(p)}_{\beta\beta} [\mathrm{yr}^{-1}] \bigg[ \frac{N_A \: 10^3 \: \eta(^{130}\mathrm{Te})} {m(\mathrm{TeO}_2)\:[\rm{g/mol}]} \bigg]
                        \epsilon_{s} \cdot \\
                    & \cdot (M\Delta t)_{ds}\:\mathrm{[kg\cdot yr]} \:
                        (\epsilon^{cut})_{ds}^M \:(\epsilon^{acc})_{ds}
\end{split}
\label{eq:LambdaS}
\end{equation}
where $\Gamma^{(p)}_{\beta\beta}$ is the decay rate of process $p$ and the other parameters were introduced following Eq. \ref{eq:SensitivityTilde}.
The $\Gamma_{\beta\beta}^{(p)}$ parameter describes the rate of the process  $p = 0\nu, 2\nu$ under investigation and is given in both cases a uniform physical prior, $\Gamma_{\beta\beta} > 0$.

The background term $f_B(E_{ev})$ is parameterized as
\begin{equation}
    f_B(E_{ev}) = \frac{1}{\Delta E} \bigg[ 1 + m_{s} \big( E_{ev} - E^{(s)}_0 \big) \bigg]
\end{equation}
where $\Delta E = E^{max}_{s} - E^{min}_{s}$ is the width of the region of interest, $E^{(s)}_0$ is the center of it, and $m_s$ describes the slope of the background for signature $s$. The normalization of the background term represents the number of expected background counts
\begin{equation}
    \lambda_{B \: s,ds} =  \mathrm{BI}_{s} \: (M\Delta t)_{ds} \: 
    ( E^{max}_{s} - E^{min}_{s} )
    \label{eq:LambdaB_UEMAP}
\end{equation}
where BI$_s$ is the background index for signature $s$.
Background simulations suggest that a uniform event distribution is enough to describe the continuous background in all signatures except the $2\nu\beta\beta$ 2A0-2B1. For this reason the $m_s$ parameter is included only when necessary.
The background is fully described by the BI$_s$ and $m_s$ which, together, make 4 (3) nuisance parameters in the $2\nu\beta\beta$ $(0\nu\beta\beta)$ case respectively, that will be marginalized over.
The prior for background indices BI$_s$ and slopes $m_s$ is uniform.

The combined log-likelihood reads
\begin{equation}
    \log \mathcal{L}(\mathcal{D}|\mathrm{H}_{S+B}) = \sum_{s,ds} \log \mathcal{L}_{s,ds}
\end{equation}
where $\mathrm{H}_{S+B}$ indicates that the likelihood is written in the signal-plus-background model hypothesis H$_{S+B}$, i.e. that the existence of the process of interest is assumed. \linebreak
The background-only hypothesis H$_B$ is a particular case that can be obtained by setting $\Gamma_{\beta\beta} = 0$.

\subsection{Blinding and Sensitivity}
\label{sec:BlindingAndSensitivity}
We blind the data by injecting simulated signal events into the experimental spectrum. 
We inject a random and unknown number of fake signal events that would correspond to an event rate larger than the current 90 \% upper limit \cite{Adams:2018yrj,Andreotti:2011in}.
Then we compute the expected number of counts in each signature, according to known efficiencies and exposures, for each dataset. 
Each generated signal event is randomly assigned a bolometer, according to its exposure within the considered dataset. Finally the projected energy of the signal event is generated according to the detector response function
$f_S(E_{ev}|Q_s)$ centered at the expected position $Q_s$ of the
monochromatic energy release in the projected energy space. 

The injection rate of the simulated signal events is comprised between: 
\begin{equation}
\Gamma^{min}_p = 1 \cdot 10^{-23} \: [\mathrm{1/yr}]
\quad \mathrm{and} \quad
\Gamma^{max}_p = 5 \cdot 10^{-23} \: [\mathrm{1/yr}]
\end{equation}

We then fit the blinded datasets to get data driven estimates of the background levels in each fitting window. These background estimates are used as inputs to our sensitivity studies in the next section.

The results of the fits to the blinded data are reported in Table~\ref{tab:Blinded_5ms1200mm}.
We see a non-null background for both the \linebreak
$\mathrm{2A0-2B1}$ and $\mathrm{2A2-2B3}$ signatures. No background is expected for the $\mathrm{3A0}$ signature.

To extract the median half-life sensitivity for each decay, we generate $10^4$ background-only \textit{Toy Monte Carlo simulations} (ToyMC), using the numbers in Tab. \ref{tab:Blinded_5ms1200mm}.
A background-only ToyMC simulation is an ensemble of simulated \linebreak
datasets, according to the following procedure which is iterated $N_{toy}$ times, to produce the same number of ToyMC ensembles.
We define a set of signatures, together with the multiplicity and cuts in the ordered energy variables that identify candidate events. For each signature, we set a functional form for the background pdf, either constant or linear, and sample a value from the posterior pdf of the corresponding blinded fit. We compute the number of expected background events for each signature and dataset according to Eq. \ref{eq:LambdaB_UEMAP} and sample the actual number of background events from a Poisson distribution with expectation value equal to the number of expected counts. 

We store each simulated ToyMC event as a vector of ordered energy releases $\vec E_{ev}$ and related bolometers $\vec{\mathrm{ch}}_{ev}$, where the bolometers are randomly extracted from the active \linebreak 
bolometers of each dataset according to their exposure in the data, while the energies are generated according to the selected shape of the background pdf computed with the parameters (e.g. background index) generated according to the posterior pdfs obtained with the  blinded fit to the data.

\begin{figure}
    \centering
    % trim={<left> <lower> <right> <upper>}
    \includegraphics[trim={0 0.5cm 0 1.4cm},clip,width=0.5\textwidth]
    {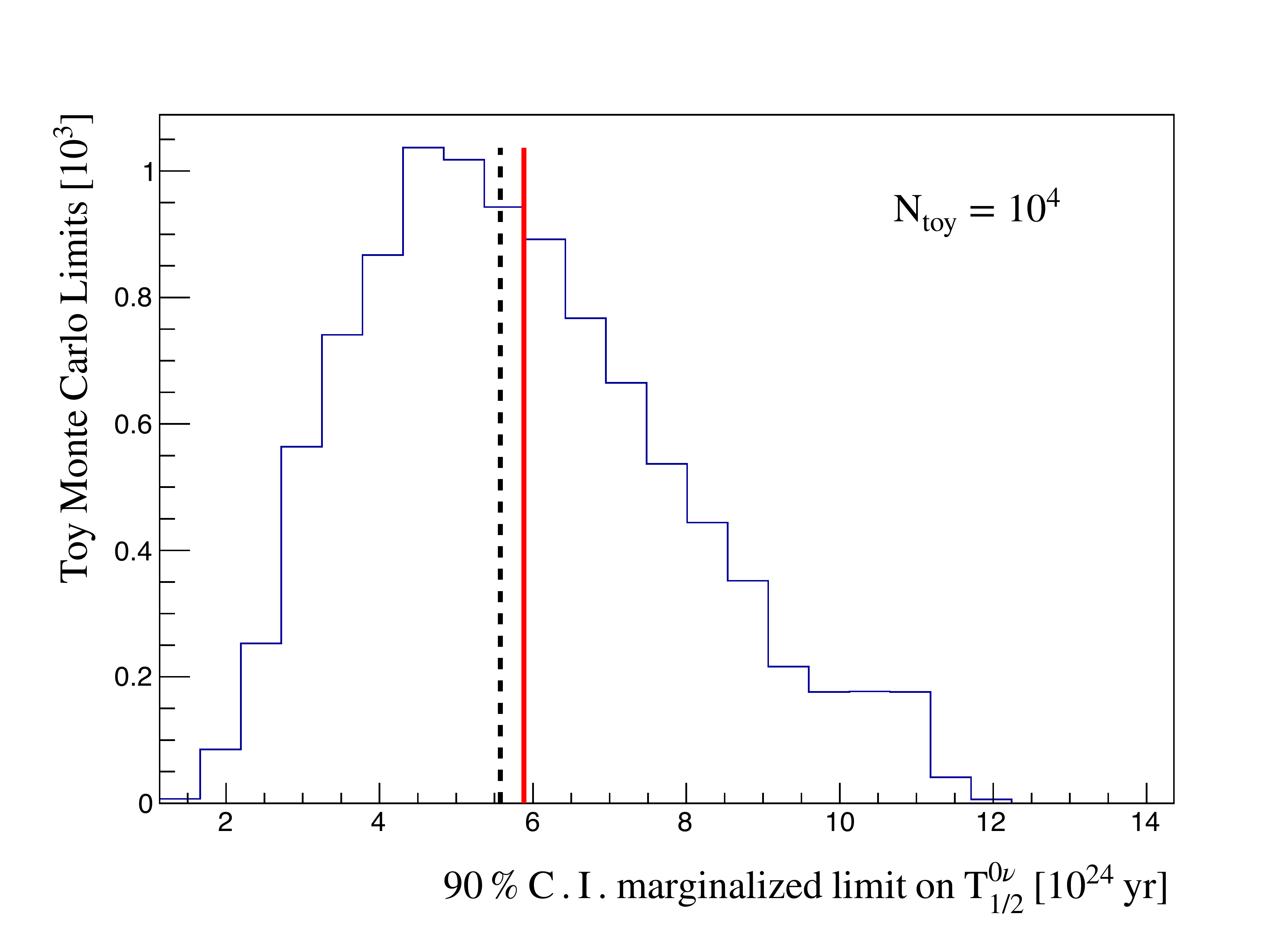}
    \includegraphics[trim={0 0.5cm 0 1.4cm},clip,width=0.5\textwidth]
    {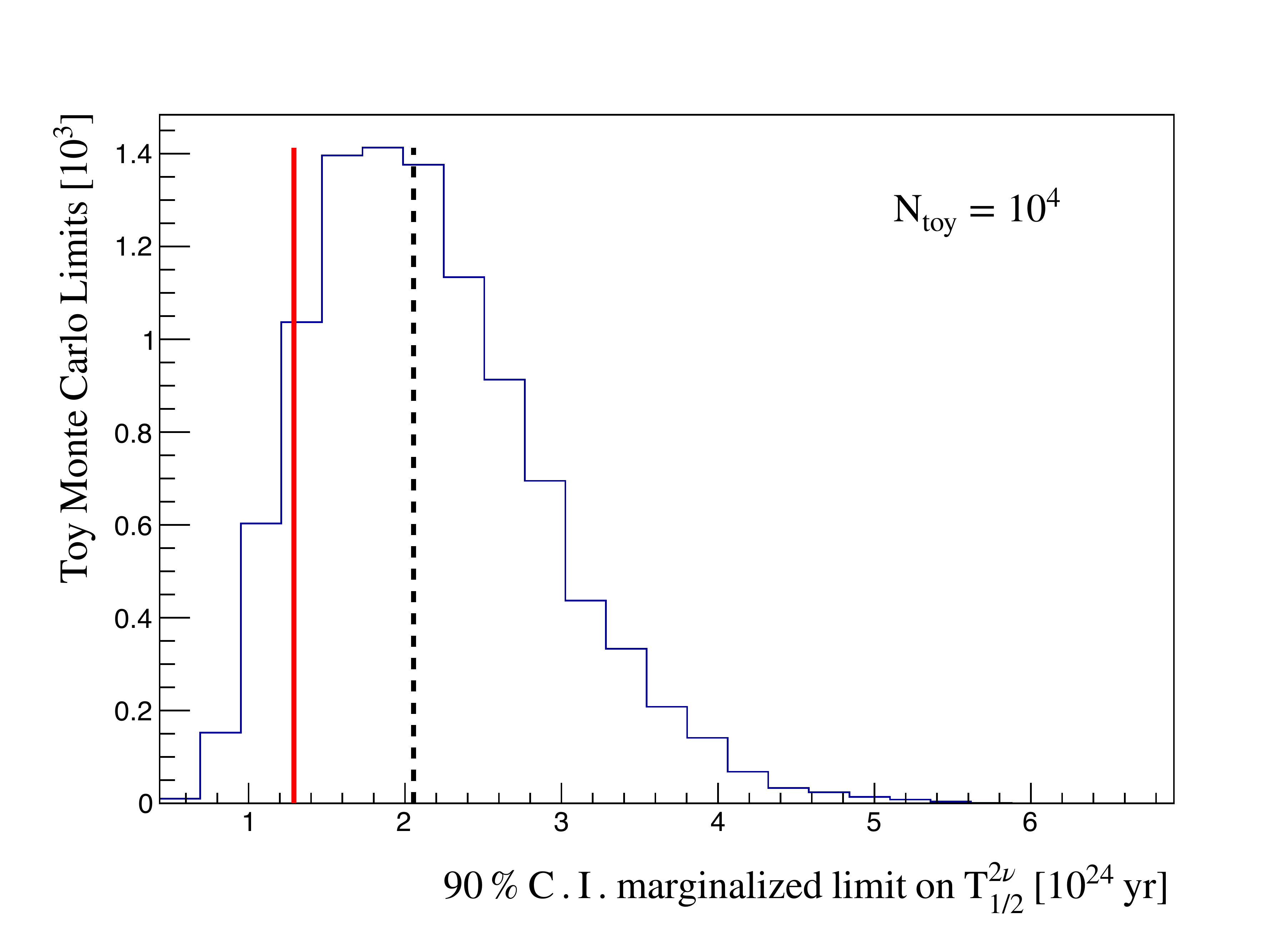}
    \caption{
    Distribution of $90 \%$ C.I. marginalized upper limits on $\mathrm{T_{1/2}} = \log 2 / \Gamma$ for $0\nu\beta\beta$ decay (top) and $2\nu\beta\beta$ decay (bottom) obtained from Toy MC simulations. 
    We obtain a median sensitivity of $S^{0\nu}_{1/2} = 2.1 \times 10^{24}$ yr and $S^{2\nu}_{1/2} = 5.6 \times 10^{24}$ yr (black dashed line), compared to the $90 \%$ C.I. limit from this analysis (red solid line).
    }
    \label{fig:LimitSettingSensitivity}
\end{figure}

We then fit each ToyMC with the signal-plus-background model $\mathrm{H}_{S+B}$ and compute the lower limit for the decay half life from the $90 \%$ quantile of the marginalized posterior pdf for the decay rate parameter. We show the distribution of such limits in Figure \ref{fig:LimitSettingSensitivity} for both the $0\nu\beta\beta$ and $2\nu\beta\beta$ decay process. We quote the half-life sensitivity as the median limit of the ToyMCs (Table \ref{tab:Blinded_5ms1200mm}). 
They are respectively: \linebreak
$\mathrm{S^{0\nu}_{1/2} = (5.6 \pm 1.4) \times 10^{24} \: \mathrm{yr}}$ for the ${0\nu}$ decay and \linebreak
$\mathrm{S^{2\nu}_{1/2} = (2.1 \pm 0.5) \times 10^{24} \: \mathrm{yr}}$ for the ${2\nu}$ decay where the uncertainty is the MAD of the corresponding distribution.

%% figure -> limit setting sensitivity w/ median for both 0nu && 2nu analyses

\begin{table}[h!]
    \centering
    
    \caption{Results of the blinded fit to $0\nu\beta\beta$ (top) and $2\nu\beta\beta$ (bottom) candidate events in the signatures of Table \ref{tab:SelectedScenarioES}. For each parameter the mean and standard deviation of the corresponding marginalized posterior distribution are reported. These values are used only as input to the sensitivity studies and fit validation. Final results reported in Table \ref{tab:Results}. }
    \label{tab:Blinded_5ms1200mm}
    
    \begin{tabular}{lcr}
    \multicolumn{3}{c}{$0\nu\beta\beta$} \\
    \hline
\textbf{Observable}                 & \textbf{Blinded Fit Value}   & \textbf{Units} \\ 
\hline
Blinded $ \mathrm{ \Gamma_{\beta\beta}^{0\nu} } $  & $ 2.8 \pm { 0.1 }  $      & $\mathrm{ 10^{ -23 }\: [yr^{-1}]}$ \\
$ \mathrm{ BI_{2A0-2B1} } $         & $ 6.1 \pm { 3.6 }  $      & $\mathrm{ 10^{ -4 }\: [counts/(keV \: kg \: yr)]}$ \\
$ \mathrm{ BI_{2A2-2B3} } $         & $ 2.7 \pm { 1.5 }  $      & $\mathrm{ 10^{ -4 }\: [counts/(keV \: kg \: yr)]}$ \\
$ \mathrm{ BI_{3A0} } $             & $ 8.7 \pm { 8.3 }  $      & $\mathrm{ 10^{ -5 }\: [counts/(keV \: kg \: yr)]}$ \\
\\
\end{tabular}
    
    \begin{tabular}{lcr}
    \multicolumn{3}{c}{$2\nu\beta\beta$} \\
    \hline
    \textbf{Observable}                 & \textbf{Blinded Fit Value}   & \textbf{Units} \\ 
    \hline
    Blinded $ \mathrm{ \Gamma_{\beta\beta}^{2\nu} } $  & $ \:5.1 \pm { 0.1 } $     & $\mathrm{ 10^{ -23 }\: [yr^{-1}]}$ \\
    $ \mathrm{ BI_{2A0-2B1} } $         & $ \:3.3 \pm { 0.4 } $     & $\mathrm{ 10^{ -3 }\: [counts/(keV \: kg \: yr)]}$ \\
    $ \mathrm{ m_{2A0-2B1} } $          & $  -5.5 \pm { 4.3 } $     & $\mathrm{ 10^{ -3 }\: [1/keV]}$ \\
    $ \mathrm{ BI_{2A2-2B3} } $         & $ \:4.0 \pm { 0.6 } $     & $\mathrm{ 10^{ -3 }\: [counts/(keV \: kg \: yr)]}$ \\
    $ \mathrm{ BI_{3A0} } $             & $ \:6.9 \pm { 6.8 } $     & $\mathrm{ 10^{ -5 }\: [counts/(keV \: kg \: yr)]}$ \\

\end{tabular}
    
\end{table}

%----------------------------------------------
\section{Results}

\begin{table}[]
    \centering
    
    \caption{
    We report here mean and standard deviation of the marginalized posterior distributions for the decay rate and background parameters for each signature, derived from unblinded data.
    The S$_{1/2}$ parameter indicates the median expected sensitivity for limit setting at $90\%$ C.I. on the T$_{1/2}$ parameter together with the MAD of its distribution.
    The last row reports the marginalized $90\%$ C.I. Bayesian lower limit on the decay half life. All results come from the combined fit with systematics. %(see Tab. \ref{tab:Systematics}).
    %$0\nu\beta\beta$ (top), $2\nu\beta\beta$ (bottom).
    }
    \label{tab:Results}
    
    \begin{tabular}{lcr}
    \multicolumn{3}{c}{$0\nu\beta\beta$} \\
    \hline
    \textbf{Parameter}          & \textbf{Final Fit Value}    & \textbf{Units} \\
    \hline
    $\mathrm{\Gamma_{\beta\beta}^{0\nu}}$    & $5.8 \pm 4.5$         & $10^{-26}$ [yr$^{-1}$] \\
    BI$_{\mathrm{2A0-2B1}}$      & $2.1 \pm 1.4$         & $10^{-4}$ [counts/(keV kg yr)] \\
    BI$_{\mathrm{2A2-2B3}}$      & $2.7 \pm 1.2$         & $10^{-4}$ [counts/(keV kg yr)] \\
    BI$_{\mathrm{3A0}}$          & $5.8 \pm 5.5$         & $10^{-5}$ [counts/(keV kg yr)] \\
    \hline
    $ \mathrm{ S^{0\nu}_{1/2} }$               & $ 5.6 \pm 1.4$              & $10^{24}$ [yr] \\
    $\mathrm{T}_{1/2}^{90\%}$   & $ > 5.9$              & $10^{24}$ [yr] \\
    \\
    \end{tabular}

    \begin{tabular}{lcr}
    \multicolumn{3}{c}{$2\nu\beta\beta$} \\
    \hline
    \textbf{Parameter}          & \textbf{Final Fit Value}    & \textbf{Units} \\
    \hline
    $\mathrm{\Gamma_{\beta\beta}^{2\nu}}$    & $2.8 \pm 1.8$         & $10^{-25}$ [yr$^{-1}$] \\
    BI$_{\mathrm{2A0-2B1}}$      & $3.0 \pm 0.3$         & $10^{-3}$ [counts/(keV kg yr)] \\
    $m_{\mathrm{2A0-2B1}}$      & $-5.2 \pm 4.2$        & $10^{-3}$  [keV$^{-1}$] \\
    BI$_{\mathrm{2A2-2B3}}$      & $4.3 \pm 0.5$         & $10^{-3}$ [counts/(keV kg yr)] \\
    BI$_{\mathrm{3A0}}$          & $5.4 \pm 5.4$         & $10^{-5}$ [counts/(keV kg yr)] \\
    \hline
    $\mathrm{ S^{2\nu}_{1/2} }$       & $ 2.1 \pm 0.5$         & $10^{24}$ [yr] \\
    $\mathrm{T}_{1/2}^{90\%}$   & $ > 1.3$              & $10^{24}$ [yr] \\
    \end{tabular}
\end{table}

\begin{figure}
    \centering
    \includegraphics[width=0.5\textwidth]{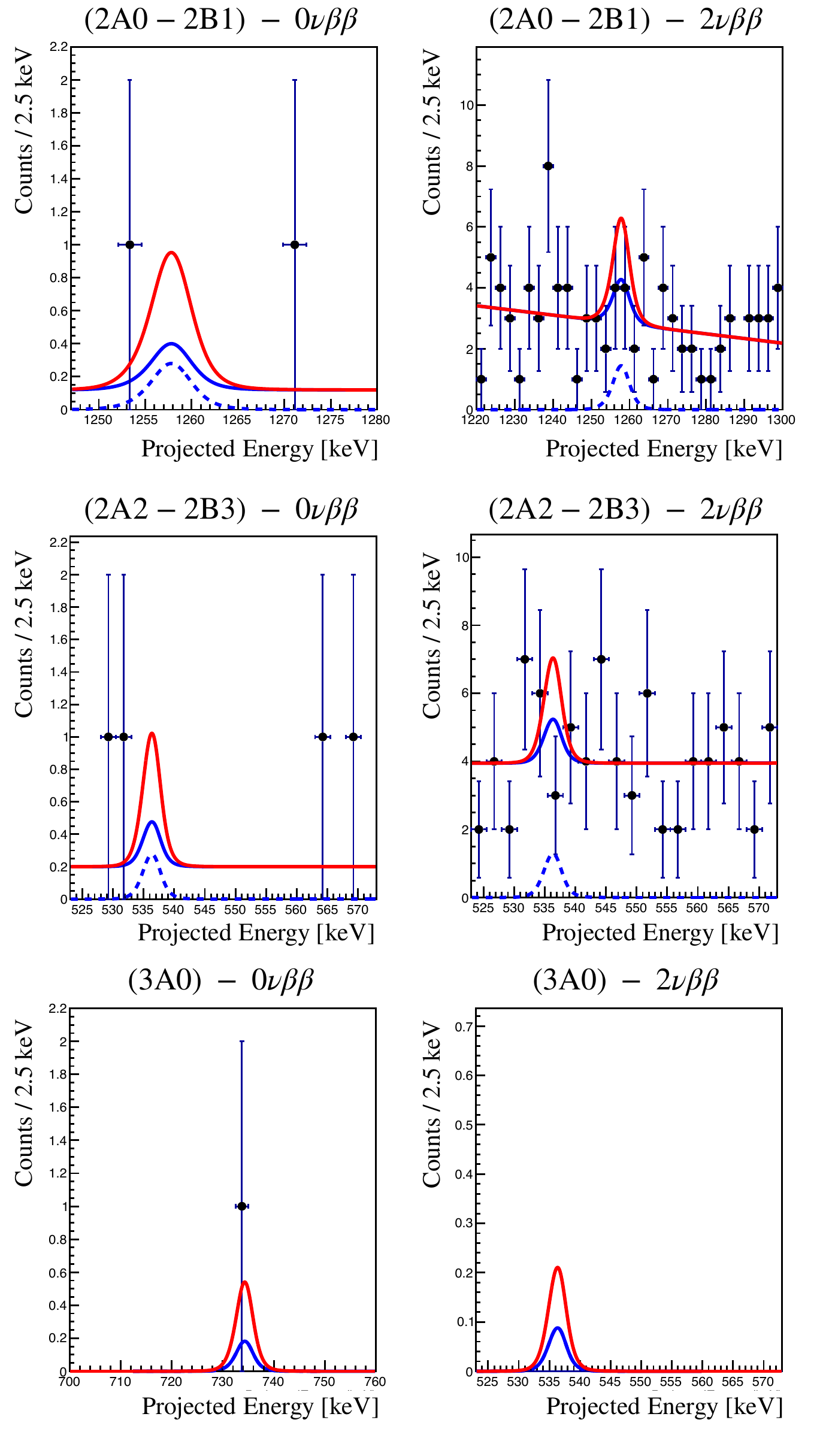}
    \caption{Result of the unbinned fit plotted on binned data. Error bars are just a visual aid and correspond to the square root of the bin contents. We show the best fit curve (blue solid), its signal component (blue dashed) at the global mode of the posterior for $0\nu\beta\beta$ (left) and $2\nu\beta\beta$ (right), and the 90 \% C.I. marginalized limit on the decay rate (red solid).}
    \label{fig:UnbinnedFit}
\end{figure}

\begin{figure}
    \centering
    \includegraphics[width=0.5\textwidth]{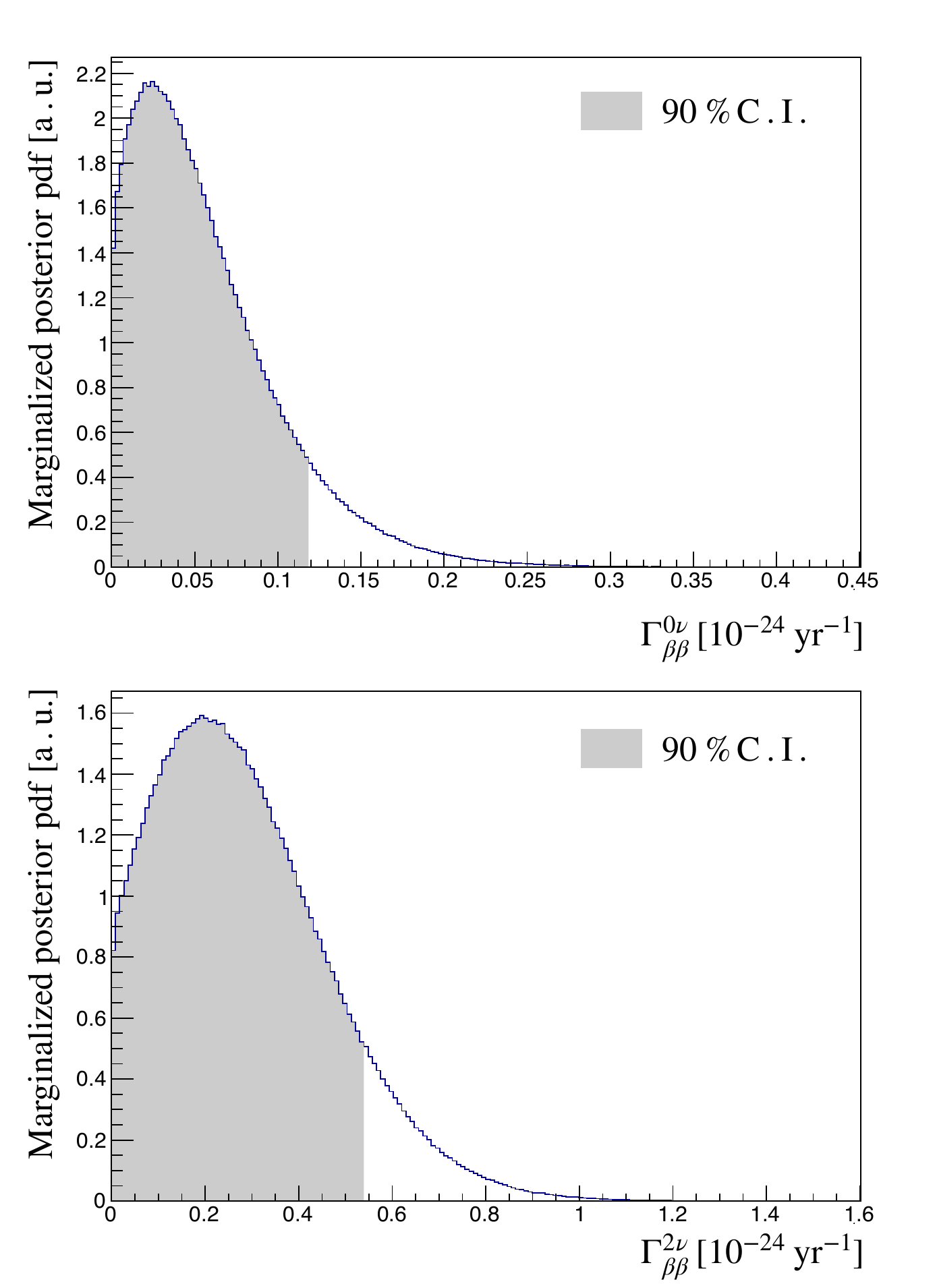}
    \caption{Marginalized decay rate posterior pdf for $0\nu\beta\beta$ (top) and $2\nu\beta\beta$ (bottom) from the combined fit with all systematics. We show the $90 \%$ C.I. in gray.}
    \label{fig:PosteriorRate}
\end{figure}

\begin{figure}[h!]
    \centering
    \includegraphics[width=0.43\textwidth]{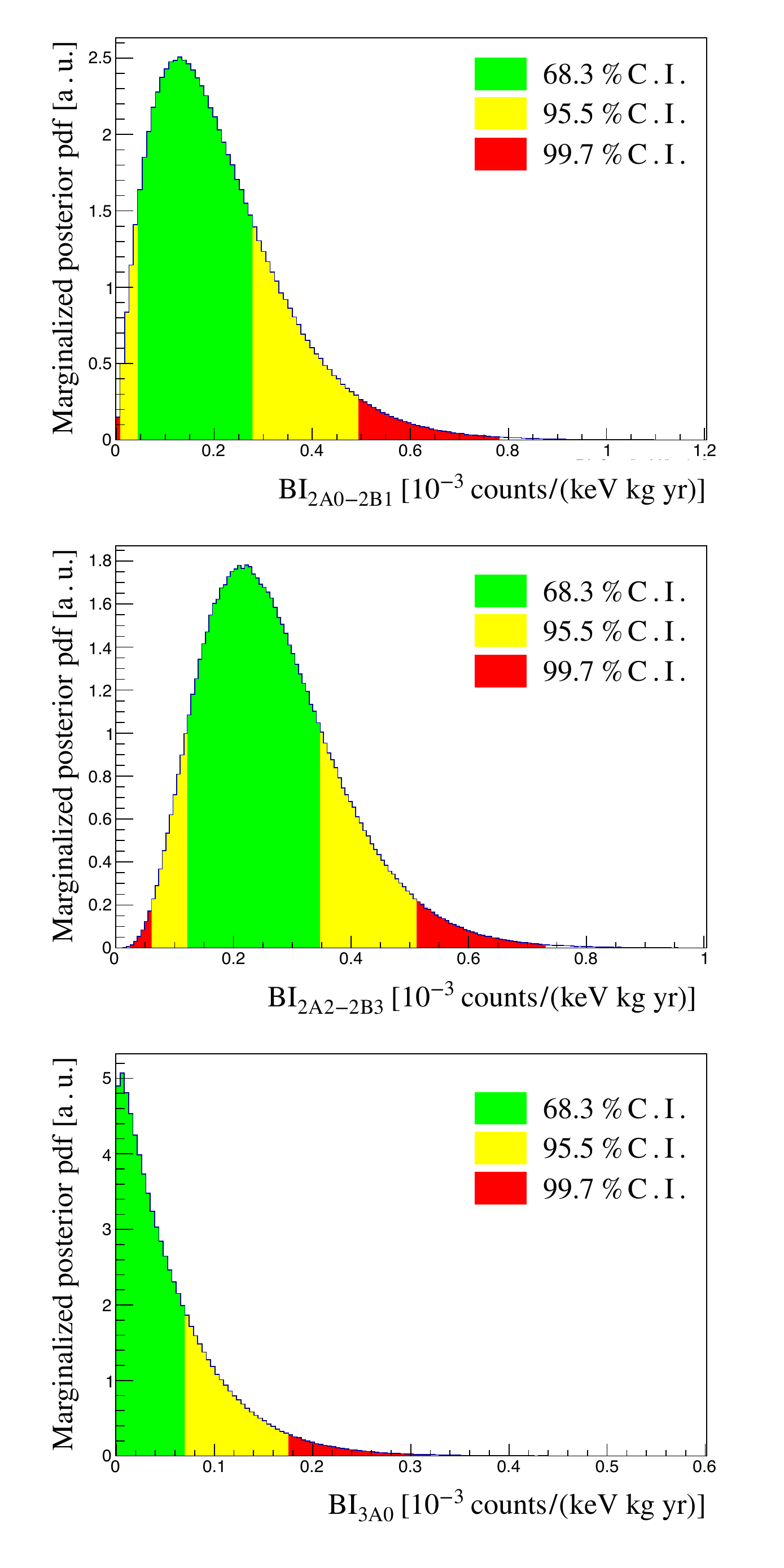}
    \caption{Marginalized posterior pdf for the background parameters of the $0\nu\beta\beta$ model from the combined fit with all systematics. We show the $68.3 \%, 95.5 \%, 99.7 \%$ smallest C.I. in green, yellow and red respectively. }
    \label{fig:PosteriorBkg0nbb}
\end{figure}

\begin{figure}[h!]
    \centering
    \includegraphics[width=0.43\textwidth]{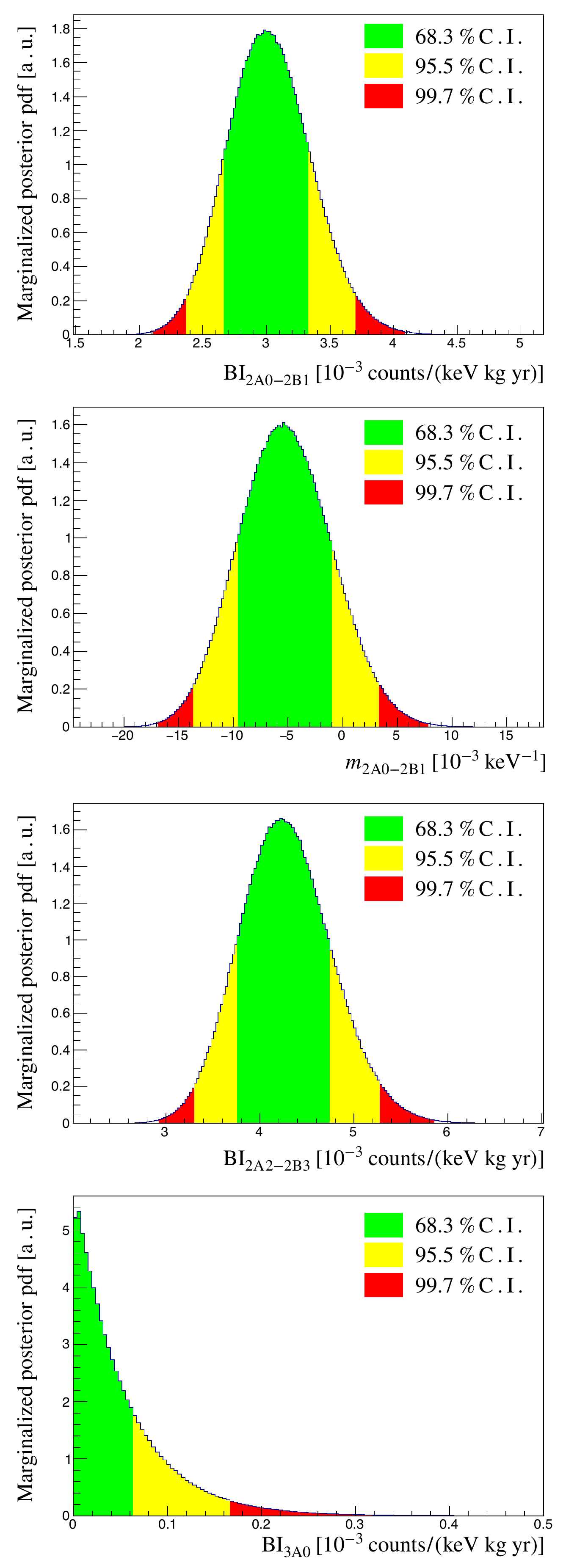}
    \caption{Marginalized posterior pdf for the background parameters of the $2\nu\beta\beta$ model from the combined fit with all systematics. We show the $68.3 \%, 95.5 \%, 99.7 \%$ smallest C.I. in green, yellow and red respectively. }
    \label{fig:PosteriorBkg2nbb}
\end{figure}

%----------------------------------------------
We show in Figure~\ref{fig:UnbinnedFit} and Table~\ref{tab:Results} the results of the fit to unblinded data for both $0\nu\beta\beta$ and $2\nu\beta\beta$.
Though the data are binned for graphical reasons, the analysis is unbinned.
Including contributions from all sources of systematic uncertainty listed in Table~\ref{tab:Systematics}, no significant signal is observed in either decay mode.
The global mode of the joint posterior pdf for the rate parameter is
\begin{subequations}
    \begin{align}
    \hat{\rm{\Gamma}}^{0\nu}_{\beta\beta} = 4.0_{-4.0}^{+3.0} \times 10^{-26} \: \rm{yr}^{-1} \label{eq:GlobalMode0nbb} \\
    \hat{\rm{\Gamma}}^{2\nu}_{\beta\beta} = 2.2_{-1.9}^{+1.7} \times 10^{-25} \: \rm{yr}^{-1} \label{eq:GlobalMode2nbb} 
    \end{align}
\end{subequations}
whereas we quote the uncertainty as the marginalized $68~\%$ smallest interval.
We report the marginalized posterior pdf and the $90\%$~C.I. in Figure~\ref{fig:PosteriorRate}. We include the marginalized posterior pdfs for individual background parameters in Figure \ref{fig:PosteriorBkg0nbb} $(0\nu\beta\beta)$ and Figure \ref{fig:PosteriorBkg2nbb} $(2\nu\beta\beta)$. Their means agree with the corresponding results from blinded data within one standard deviation.
We observe a slight negative background fluctuation (i.e. limit stronger than expected) in the $0\nu\beta\beta$ decay analysis and a positive one (i.e. limit looser than expected) in the $2\nu\beta\beta$ decay analysis with respect to the median $90\%$~C.I. limit. The probability of setting an even \linebreak stronger (looser) limit in the $0\nu\beta\beta$ ($2\nu\beta\beta$) decay analysis is respectively $45.1\%$ and $10.4\%$.
Null signal rates are included in the $1\sigma$ ($2\sigma$) smallest C.I.\footnote{We refer here to the Gaussian case to define the probability content of any $n\sigma$ interval.} of the marginalized pdf for the $\rm{\Gamma}^{0\nu}_{\beta\beta}$ ($\rm{\Gamma}^{2\nu}_{\beta\beta}$) rate parameter respectively.
The following Bayesian lower bounds on the corresponding half life parameters are set: 
\begin{subequations}
    \begin{align}
    \big( \mathrm{T_{1/2}} \big)^{0\nu}_{0^+_2} &> 5.9 \times 10^{24} \: \mathrm{yr} \quad (\mathrm{90 \% \: \: C.I.}) \label{res1} \\
    \big( \mathrm{T_{1/2}} \big)^{2\nu}_{0^+_2} &> 1.3 \times 10^{24} \: \mathrm{yr} \quad (\mathrm{90 \% \: \: C.I.}) \label{res2}
    \end{align}
\end{subequations}

The results reported in this Article represent the most stringent limits on the DBD of $^{130}$Te to excited states and improve by a factor $\sim5$ the previous results on this process.

%----------------------------------------------
\subsection{Systematic Uncertainties}
\label{sec:SystematicUncertainties}
%----------------------------------------------
The major sources of systematic uncertainty are: signal efficiencies, the detector response function, energy calibration, and the uncertainty of the isotopic abundance of $^{130}$Te (see Table \ref{tab:Systematics}).
Each systematic uncertainty can be introduced as a set of nuisance parameters in the fit with a specified prior distribution. 
Each set of nuisance parameters can be activated independently with the priors listed in Table \ref{tab:Systematics}. 
We individually monitor the effect of activating each source of systematic uncertainty repeating the fit and comparing the 90\% C.I. Bayesian limit on the half life with respect to the \textit{minimal model}, where we describe all sources of systematics with constants rather than fit parameters. Finally, we repeat the fit activating all additional nuisance parameters at once. 

We include, for each dataset, two separate parameters to model different sources of uncertainty in the cut efficiency evaluation, and replace the $\varepsilon^{cut}$ constant with the sum of $\varepsilon^{cut\:(I)} + \varepsilon^{cut \: (II)}$. We refer to \textit{cut efficiency I} to parameterize the uncertainty due to the finiteness of the samples of pulser events and $\gamma$ decays used to extract the cut efficiency. Its prior is Gaussian with mean equal to $\varepsilon^{cut}$ and width equal to the corresponding uncertainty (see Table. \ref{tab:Efficiency}). The additional \textit{cut efficiency II} is uniformly distributed, with zero mean. It models the systematic uncertainty due to the PSA efficiency, shifting the cut efficiency by at most $0.7\%$.
The accidentals efficiency contributes to the systematic uncertainty just with the uncertainty due to limited statistics in the $^{40}$K peak. We add one  nuisance parameter per dataset with a Gaussian distributed prior to model this effect.
The containment efficiency is instead affected by uncertainty due to the simulation of Compton scattering events at low energy. The uncertainty due to the number of simulated signal events is negligible. We take the ratio of the Compton scattering attenuation coefficient to reference data \cite{Allison:2016lfl} as a measure of the relative uncertainty on this efficiency term. We account for this, for each signature, introducing a nuisance containment efficiency parameter with a Gaussian prior.
The $^{130}$Te natural isotopic abundance on natural Te is modeled with a single global nuisance parameter with a Gaussian prior $\eta = (34.167 \pm 0.002) \%$.
Both the detector response function shape and the energy scale bias are evaluated from data, as anticipated in Sec. \ref{sec:fitting}. Each effect is separately parameterized with a 2nd order polynomial as a function of energy, whose coefficients are evaluated with a fit to the 5-7 most visible peaks in each dataset. The uncertainty and correlations among such parameters are themselves a source of systematic uncertainty, and are included as 2 independent sets of correlated parameters per dataset, with a multivariate normal prior distribution.
In this way, the detector response function width and position are allowed to float within their uncertainty.

Uncertainty in modeling the detector response function leads to the dominant systematic effect on the limit, which is below a 1\% shift. Sub-dominant effects come from the energy scale bias and the containment efficiency (Table \ref{tab:Systematics}). 

\begin{table}[]
    \centering
    \begin{tabular}{llcc}
    \multirow{2}{*}{Source}                  & \multirow{2}{*}{Prior}         & \multicolumn{2}{c}{Effect on $\mathrm{T}_{1/2}^{90\%}$} \\ 
                            &               & $0\nu\beta\beta$ & $2\nu\beta\beta$ \\
    \hline
    Cut efficiency I        & Gaussian      & $0.2 \%$ & $<0.1\%$\\
    Cut efficiency II       & Uniform       & $0.1\% $ & $0.1\%$\\
    Accidentals efficiency  & Gaussian      & $0.2\% $ & $<0.1\%$\\
    Containment efficiency  & Gaussian      & $0.3\% $ & $0.1\%$\\
    $^{130}\mathrm{Te}$ isotopic abundance 
                            & Gaussian      & $<0.1\%$ & $0.1\%$\\
    Energy scale bias       & Multiv. Gaussian  & $0.3\% $ & $0.1\%$\\
    Detector resolution     & Multiv. Gaussian  & $0.5\% $ & $0.4\%$\\
    \hline
    Combined                & Multivariate  & $0.4\% $ & $0.1\%$\\
    \end{tabular}
    \caption{
    Systematic uncertainties. We report each effect separately and their combination on the marginalized $90 \%$ C.I. T$_{1/2}^{90\%}$ limit on the $0\nu\beta\beta$ and $2\nu\beta\beta$ decay half life. 
    }
    \label{tab:Systematics}
\end{table}

\section{Conclusions}

We have presented the latest search for DBD of $^{130}$Te to the first $0^+$ excited state of $^{130}$Xe with CUORE based on a 372.5 kg$\cdot$yr TeO$_2$ exposure. We found no evidence for either $0\nu\beta\beta$ nor $2\nu\beta\beta$ decay and we placed a Bayesian lower bound at $90\%$ C.I. on the decay half lives of \linebreak
$\mathrm{(T_{1/2})^{0\nu}_{0^+_2} > 5.9 \times 10^{24} \: \mathrm{yr}}$ 
for the $0\nu\beta\beta$ mode and \linebreak
$\mathrm{(T_{1/2})^{2\nu}_{0^+_2} > 1.3 \times 10^{24} \: \mathrm{yr}}$
for the $2\nu\beta\beta$ mode.

The median limit setting sensitivity for the $2\nu\beta\beta$ decay of $2.1 \times 10^{24}$yr is starting to approach the $7.2 \times 10^{24}$yr lower bound of the QRPA theoretical predictions half life range for this decay mode. The CUORE experiment is steadily taking data at an average rate of $\sim 50$~kg$\cdot$yr/month and by the end of its data taking the collected exposure is expected to increase by an order of magnitude. 
Work is ongoing to improve the sensitivity by extending the analysis to not-fully-contained events, leveraging the information from the topology of higher dimension coincident signal multiplets to further reduce the background, and improving the signal efficiency by lowering the threshold of the pulse shape discrimination algorithm.

\begin{acknowledgements}
The CUORE Collaboration thanks the directors and staff of the Laboratori Nazionali del Gran Sasso and the technical staff of our laboratories. 
This work was supported by the Istituto Nazionale di Fisica Nucleare (INFN); the National Science Foundation under Grant Nos. NSF-PHY-0605119, NSF-PHY-0500337, NSF-PHY-0855314, NSF-PHY-0902171, NSF-PHY-0969852, NSF-PHY-1307204, NSF-PHY-1314881, NSF-PHY-1401832, and NSF-PHY-1913374; and Yale University. 
This material is also based upon work supported by the US Department of Energy (DOE) Office of Science under Contract Nos. DE-AC02-05CH11231 and DE-AC52-07NA27344; by the DOE Office of Science, Office of Nuclear Physics under Contract Nos. DE-FG02-08ER41551, DE-FG03-00ER41138, DE-SC0012654, DE-SC0020423, DE-SC0019316; and by the EU Horizon2020 research and innovation program under the Marie Sklodowska-Curie Grant Agreement No. 754496.  
This research used resources of the National Energy Research Scientific Computing Center (NERSC).
This work makes use of both the DIANA data analysis and APOLLO data acquisition software packages, which were developed by the CUORICINO, CUORE, LUCIFER and CUPID-0 Collaborations.
We acknowledge contributions from J. Kotila on dedicated computation of beta spectra.
\end{acknowledgements}

\bibliographystyle{unsrt}
\bibliography{ref}

\begin{comment}
\clearpage
\newpage
\pagebreak[4]
\end{comment}

\end{document}